\documentclass[12pt,preprint]{aastex}
\usepackage{graphicx}

\lefthead{JUNG ET AL.}
\righthead{ }

\begin{document}
\title{
OGLE-2017-BLG-1522: A giant planet around a brown dwarf located in the Galactic bulge
}

\author{
Y.~K.~Jung$^{1,10}$, 
A.~Udalski$^{2,11}$,
A.~Gould$^{1,3,4,10}$,
Y.-H.~Ryu$^{1,10}$, 
J.~C.~Yee$^{5,10}$ \\
and \\
C.~Han$^{6}$, M.~D.~Albrow$^{7}$, C.-U.~Lee$^{1,8}$, S.-L.~Kim$^{1,8}$, 
K.-H.~Hwang$^{1}$, S.-J.~Chung$^{1,8}$, I.-G.~Shin$^{5}$, W.~Zhu$^{3}$, 
S.-M.~Cha$^{1,9}$, D.-J.~Kim$^{1}$, Y.~Lee$^{1,9}$, B.-G.~Park$^{1,8}$, 
D.-J.~Lee$^{1}$, H.-W.~Kim$^{1}$, R.~W.~Pogge$^{4}$ \\ 
(The KMTNet Collaboration) \\
M.~K.~Szyma{\'n}ski$^{2}$, P.~Mr{\'o}z$^{2}$, R.~Poleski$^{2,3}$, J.~Skowron$^{2}$, 
P.~Pietrukowicz$^{2}$, I.~Soszy{\'n}ski$^{2}$,S.~Koz{\l}owski$^{2}$, K.~Ulaczyk$^{2}$, 
M.~Pawlak$^{2}$, K.~Rybicki$^{2}$ \\
(The OGLE Collaboration)
}

\bigskip\bigskip
\affil{$^{1}$Korea Astronomy and Space Science Institute, Daejon 34055, Republic of Korea}
\affil{$^{2}$Warsaw University Observatory, Al. Ujazdowskie 4, 00-478 Warszawa, Poland}
\affil{$^{3}$Department of Astronomy, Ohio State University, 140 W. 18th Ave., Columbus, OH 43210, USA}
\affil{$^{4}$Max-Planck-Institute for Astronomy, K$\rm \ddot{o}$nigstuhl 17, 69117 Heidelberg, Germany}
\affil{$^{5}$Harvard-Smithsonian Center for Astrophysics, 60 Garden St., Cambridge, MA 02138, USA}
\affil{$^{6}$Department of Physics, Chungbuk National University, Cheongju 28644, Republic of Korea}
\affil{$^{7}$University of Canterbury, Department of Physics and Astronomy, Private Bag 4800, Christchurch 8020, New Zealand}
\affil{$^{8}$Korea University of Science and Technology, 217 Gajeong-ro, Yuseong-gu, Daejeon 34113, Korea}
\affil{$^{9}$School of Space Research, Kyung Hee University, Yongin 17104, Republic of Korea}
\footnotetext[10]{The KMTNet Collaboration.}
\footnotetext[11]{The OGLE Collaboration.}

\begin{abstract}
We report the discovery of a giant planet in the OGLE-2017-BLG-1522 microlensing 
event. The planetary perturbations were clearly identified 
by high-cadence survey experiments despite the relatively short event timescale of 
$t_{\rm E} \sim 7.5$ days. The Einstein radius is unusually small, 
$\theta_{\rm E} = 0.065\,$mas, implying that the lens system either has 
very low mass or lies much closer to the microlensed source than the Sun, or both. 
A Bayesian analysis yields component masses
$(M_{\rm host}, M_{\rm planet})=(46_{-25}^{+79}, 0.75_{-0.40}^{+1.26})~M_{\rm J}$
and source-lens distance $D_{\rm LS} = 0.99_{-0.54}^{+0.91}~{\rm kpc}$, implying that this 
is a brown-dwarf/Jupiter system that probably lies in the Galactic bulge,
a location that is also consistent with the relatively low lens-source
relative proper motion $\mu= 3.2 \pm 0.5~{\rm mas}~{\rm yr^{-1}}$. 
The projected companion-host separation is 
$0.59_{-0.11}^{+0.12}~{\rm AU}$, indicating that the planet is placed 
beyond the snow line of the host, i.e., $a_{sl} \sim 0.12~{\rm AU}$. 
Planet formation scenarios combined with the small companion-host mass ratio $q \sim 0.016$ 
and separation suggest that the companion could be the first discovery of a giant planet 
that formed in a protoplanetary disk around a brown dwarf host.
\end{abstract}

\section{Introduction}

Two decades after the first discoveries by \citet{wolszczan92} and \citet{mayor95}, 
about $4,000$ extrasolar planets have been discovered 
\footnote{NASA Exoplanet Archive, http://exoplanetarchive.ipac.caltech.edu}. 
Most of these planets were detected by using the radial-velocity 
(e.g., \citealt{pepe11}) or transit (e.g., \citealt{tenenbaum14}) methods. 
Planets by these methods are detected indirectly from the observations 
of their host stars, and the hosts of the planets are overwhelmingly 
nearby Sun-like stars, with a mean and standard deviation of logarithmic
mass of $\langle M_{\rm host,RV,trans}/M_\odot\rangle = -0.027 \pm 0.148$.

On the other hand, the most common population of stars in the Milky Way Galaxy 
is low-mass M dwarfs, which are difficult to observe using the radial-velocity 
or transit methods \footnote{The upcoming space missions such as the 
{\it Transiting Exoplanet Survey Satellite} \citep[TESS:][]{ricker14} will find 
transiting planets around M dwarf stars.}. 
Our galaxy could also be teeming with brown dwarfs which 
cannot sustain hydrogen fusion and thus are very faint. It has been known that these 
very low-mass (VLM) objects can have circumstellar disks, which are believed to be 
formed by the process of grain growth, through dust settling, followed by 
crystallization \citep{apai05,riaz12}. If these disks can provide 
enough materials, then, planets can be formed in the disks of such low luminosity 
objects \citep{luhman12}.

Studying planets around VLM objects are important not only because these objects are 
common but also because it can help us to better understand the planet formation mechanism. 
For example, the core accretion theory predicts that giant planets are difficult to be formed 
in the disks of low-mass stars \citep{ida04,laughlin04,kennedy06}, hence the formation rate of 
giant planets in VLM objects should be low. By contrast, according to the disk instability 
mechanism \citep{boss06}, giant planets can form in abundance and thus the giant-planet formation 
rate would be high. Therefore, determination of the planet formation rate based on a large 
unbiased sample of VLM objects can provide us with an important constraint on the planet 
formation mechanism.

Despite their usefulness, few planets orbiting VLM objects are known to date 
due to the observational difficulty caused by the faintness of planet hosts. 
Although some planetary systems were discovered by the direct imaging method 
\citep{chauvin04,todorov10,gauza15,stone16}, the sample is greatly biased toward 
planets with very wide separations $(\gg 1 {\rm AU})$ from their hosts. Furthermore, 
it is difficult to spectroscopically determine the masses of the planets 
due to the faintness of the host combined with the extremely 
long orbital periods \citep{joergens07}.

Microlensing can provide a complementary channel to detect and characterize planets 
around VLM objects. Since lensing effects occur solely by the gravity of a lensing object, 
the microlensing method is suitable to detect planets around faint or even dark VLM objects, 
implying no bias by the brightness of the planet hosts. Furthermore, the method is sensitive 
to planets in a wide separation range of $ \sim 0.2 - 10~{\rm AU}$, which cover the region 
of giant planet formation. The method already proved its usefulness by detecting planets 
around very low-mass host stars or brown dwarfs 
(e.g., \citealt{street13,furusawa13,han13,skowron15,shvartzvald17,nagakane17}).

In this paper, we report the discovery of a planet orbiting a brown dwarf from 
the analysis of the OGLE-2017-BLG-1522 microlensing event. Although the event timescale 
is relatively short, the anomaly is clearly captured from continuous observations 
of the Korea Microlensing Telescope Network \citep[KMTNet:][]{kim16} survey.

\section{Observation}

The OGLE-2017-BLG-1522 event occurred at 
$(\alpha,\delta)_{\rm J2000} = (18^{\rm h}01^{\rm m}16^{\rm s}\hskip-2pt.65, 
-28^{\circ}27'43''\hskip-2pt.1)$ $[(l,b)=(2.151^\circ, -2.179^\circ)]$. 
It was discovered on 7 August 2017 by the Optical Gravitational Lensing 
Experiment \citep[OGLE:][]{udalski15} using the 1.3m Warsaw telescope 
at the Las Campanas Observatory in Chile, and announced from the Early 
Warning System \citep[EWS:][]{udalski03}

The lensed star was also monitored by KMTNet. The KMTNet 
observations were carried out using three 1.6m telescopes located at 
the Cerro Tololo Inter-American Observatory in Chile (KMTC), 
South African Astronomical Observatory in South Africa (KMTS), and 
Siding Spring Observatory in Australia (KMTA). The event lies in 
one of the pairs of its two offset fields (BLG03 and BLG43) with 
combined $4~{\rm hr}^{-1}$ observation cadence. From this feature and
with telescopes that are globally distributed, the KMTNet survey 
continuously and densely covered the event.

Data reductions were processed using pipelines of the individual 
groups \citep{udalski03, albrow09}, which are based on the difference 
image analysis \citep[DIA:][]{alard98}. For the usage of the data sets 
obtained from different observatories and reduced from different pipelines, 
we renormalized the errors of each data set. Following the procedure 
described in \citet{yee12}, we adjusted the errors as 
\begin{equation}
\sigma^\prime = \sqrt{\sigma_{\rm min}^2 + (k\sigma_{0})^2}.
\label{eq1}
\end{equation}
where $\sigma_{0}$ is the error determined from the pipeline and 
$\sigma_{\rm min}$ and $k$ are the adjustment parameters. We note that 
the errors of OGLE data set were adjusted using the prescription discussed 
in \citet{skowron16}. In Table~\ref{table:one}, we present the correction 
parameters of individual data sets with the observed passbands and 
the number of data.

\section{Analysis}

\begin{figure*}[th]
\epsscale{0.7}
\plotone{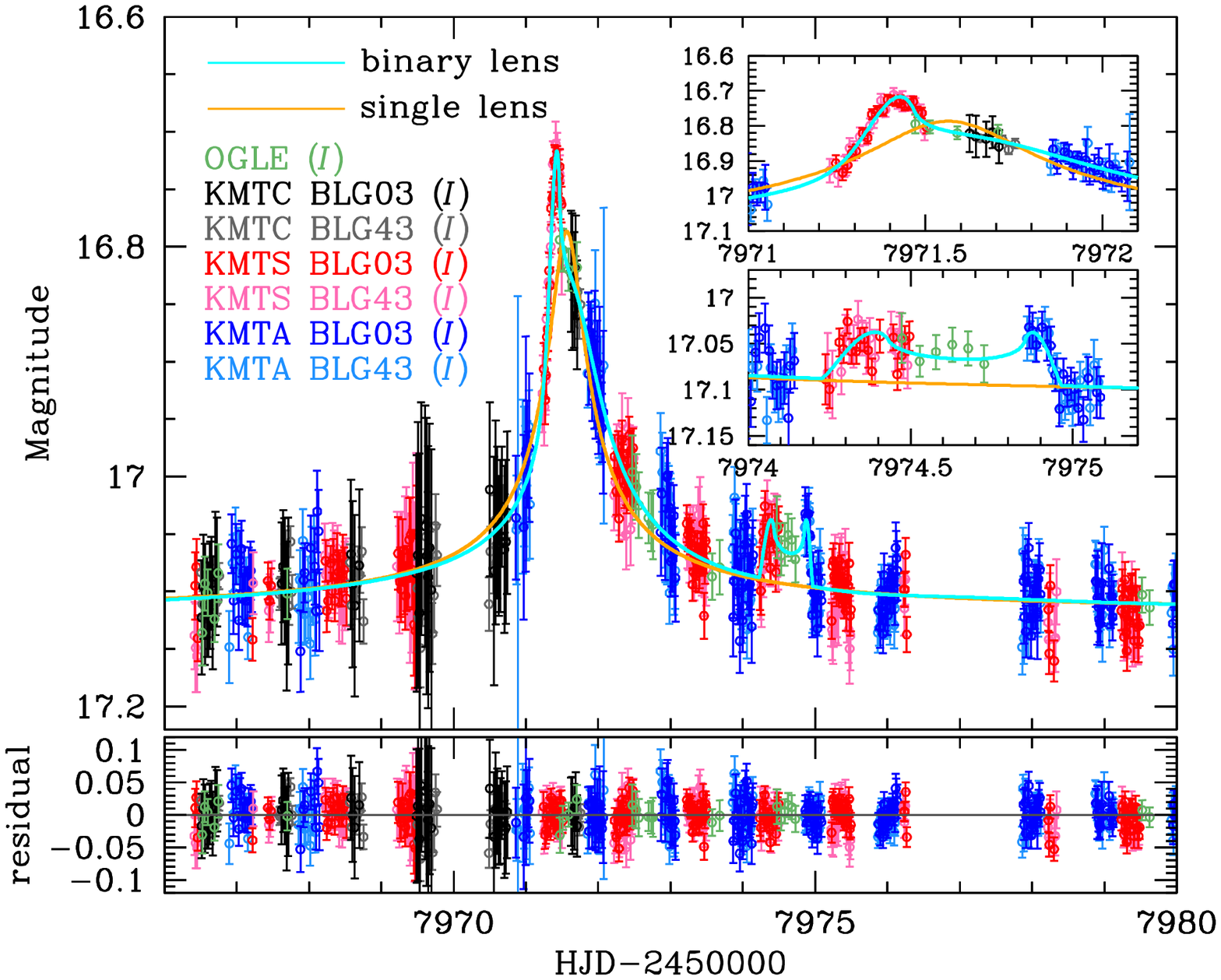}
\caption{\label{fig:one}
Light curve of OGLE-2017-BLG-1522. The upper and lower insets show the 
zoom of two major perturbations centered at ${\rm HJD}' \sim 7971.4$ and $7974.6$, 
respectively. The cyan curve is the model curve based on the binary-lens 
interpretation, while the orange curve is based on the single-lens interpretation.
}
\end{figure*}

As presented in Figure~\ref{fig:one}, the OGLE-2017-BLG-1522 light curve shows 
deviations from a standard single-mass lensing curve \citep{paczynski86}. 
The deviations are composed of two major perturbations, one strong perturbation 
at ${\rm HJD}'(={\rm HJD} - 2,450,000~{\rm days}) \sim 7971.4$ and the other weak short-term 
perturbation in the region between $7974 < {\rm HJD}' < 7975$. The latter anomaly 
consists of a trough centered at ${\rm HJD}' \sim 7974.6$ surrounded by bumps 
at both sides. Such a short-term anomaly is a characteristic feature that occurs 
when the source star crosses the small caustic induced by a low-mass companion to 
the primary lens. Therefore, we examine the event with the binary-lens interpretation.

\begin{figure}[th]
\epsscale{0.7}
\plotone{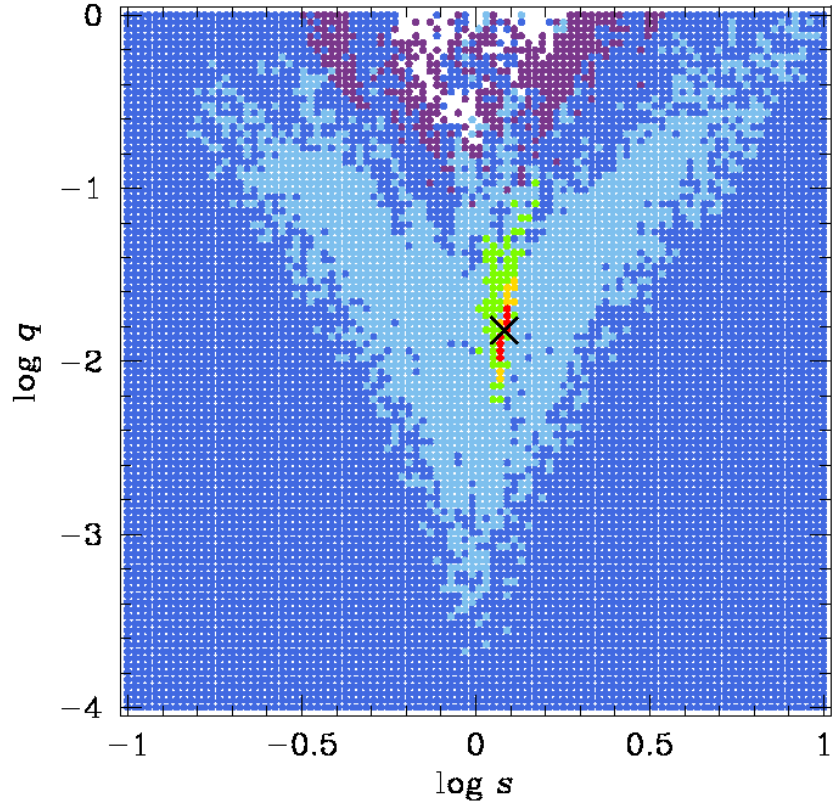}
\caption{\label{fig:two}
$\Delta\chi^{2}$ surface in $(s, q)$ space obtained from the grid search. 
The space is color coded by $\Delta\chi^2$ $< 5^{2}$ (red), $< 10^{2}$ (yellow), 
$< 15^{2}$ (green), $< 20^{2}$ (light blue), $< 25^{2}$ (blue), 
and $< 30^{2}$ (purple) level, respectively. 
The cross mark is the location of the best-fit solution.     
}
\end{figure}

To find the lensing parameters that describe the light curve, 
we adopt the parametrization and follow the procedure presented in \citet{jung15}. 
We first carry out a preliminary grid search by setting $s$, $q$, and $\alpha$ as 
independent variables. Here $s$ and $q$ are, respectively, the projected separation 
(normalized to the angular Einstein radius, $\theta_{\rm E}$) 
and the mass ratio of the binary lens, and $\alpha$ is the trajectory angle. 
The grid space $(s, q, \alpha)$ is divided into $(100, 100, 21)$ grids and the ranges of 
individual variables are $-1.0 < {\rm log}~s < 1.0$, $-4.0 < {\rm log}~q < 0.0$, and 
$0 < \alpha < 2\pi$, respectively. We note that $(s,q)$ are fixed, while $\alpha$ 
is allowed to vary at each grid point. From this search, we identify only one local minimum. 
It can be seen in Figure~\ref{fig:two}, where we show the derived $\Delta\chi^2$ surface 
in the $(s, q)$ space. We then investigate the local solutions and find the global 
minimum by optimizing all parameters including grid variables 
with the Markov Chain Monte Carlo (MCMC) method.

\begin{figure}[th]
\epsscale{0.7}
\plotone{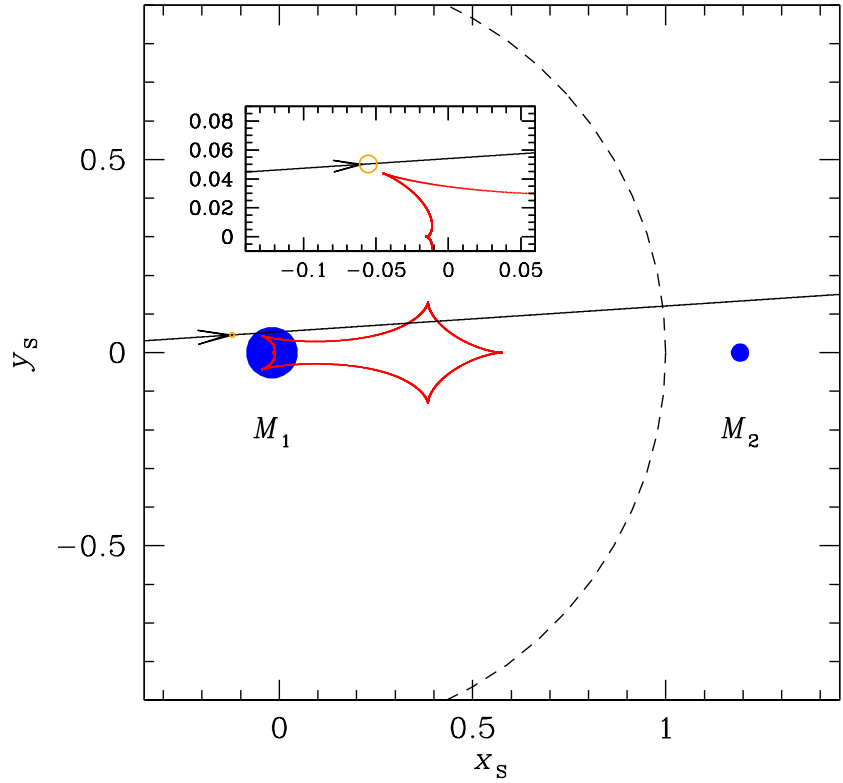}
\caption{\label{fig:three}
Geometry of the best-fit solution. The red closed curve is the caustic, 
the two blue dots ($M_{1}$ and $M_{2}$) are the positions of 
binary-lens components, and the orange circle is the size of the source. 
The dashed circle is the angular Einstein ring of the lens system and all 
lengths are scaled to its radius $\theta_{\rm E}$. 
}
\end{figure}

In Table~\ref{table:two}, we list the best-fit solution determined from our modeling. 
The model curve of the solution is presented in Figure~\ref{fig:one}. Also shown in 
Figure~\ref{fig:three} is the corresponding lensing geometry. We find that the 
companion-host separation is $s = 1.21$ and the companion-host mass 
ratio is $q \sim 0.016$, implying that the companion would be in 
the planetary or substellar regime. In this case, the binary lens induces 
a single 6-sided resonant caustic near the host star (e.g., \citealt{dominik99}). 
Although this caustic is resonant, it is very close to separating 
into a 4-sided ``central caustic'' (left) and a 4-sided ``planetary caustic'' (right), 
which it would do if $s$ were only slightly larger than its actual value of $s = 1.21$. 
Specifically the transition would occur at \citep{erdl93}
\begin{equation}
\label{eq2}
s_{\rm transition} = \sqrt{(1 + q^{1/3})^3\over 1+q} \rightarrow 1.39.
\end{equation}
Hence, the cusps associated with the central-caustic wing of resonant
caustic are strong, while those associated with the planetary-caustic wing
are weak. The two major perturbations were generated by the source transit 
through the caustic. The first perturbation occurred when the source 
passed close to one of the central-caustic-wing cusps, while the other perturbation 
occurred when the source transited the caustic near one of the planetary-caustic-wing cusps.

\begin{figure}[th]
\epsscale{0.7}
\plotone{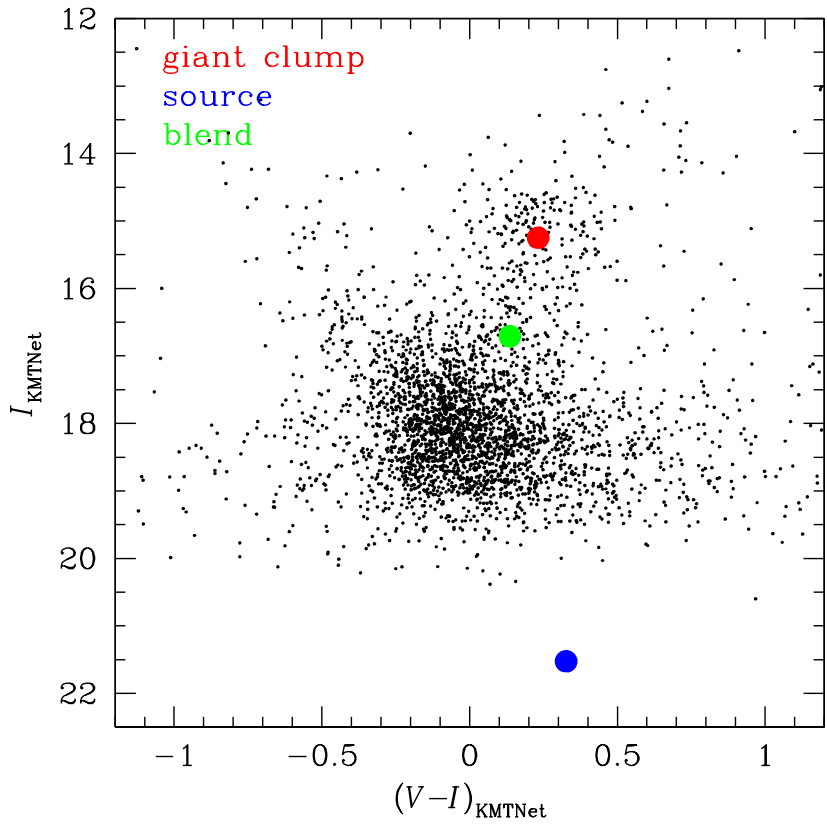}
\caption{\label{fig:four}
Position of the source star (blue dot) relative to the GC centroid (red dot) 
in the $(V-I, I)$ CMD of field stars around OGLE-2017-BLG-1522. 
The green dot is the position of blended light. 
}
\end{figure}

We clearly detect finite-source effects from which the normalized source
radius $\rho_{*}$ and the angular Einstein radius $\theta_{\rm E}$ is determined by 
\begin{equation}
\theta_{\rm E} = {\theta_{*} \over \rho_{*}}.
\label{eq3}
\end{equation}
Here the angular size of the source $\theta_{*}$ is estimated from the calibrated 
brightness $I_{0,\rm S}$ and color $(V-I)_{0,\rm S}$ of the source. To find 
$(V-I, I)_{0,\rm S}$, we follow the procedure of \citet{yoo04}. First, we build 
the instrumental color-magnitude diagram (CMD) using the KMTC Dophot reductions 
(Figure~\ref{fig:four}). Next, we estimate $(V-I, I)_{0,\rm S}$ with the equation
\begin{equation}
(V-I, I)_{0,\rm S} = (V-I, I)_{\rm S} - (V-I, I)_{\rm GC} + (V-I, I)_{0,\rm GC},
\label{eq4}
\end{equation}
where $(V-I, I)_{\rm S}$ and $(V-I, I)_{\rm GC}$ are, respectively, the positions of the source 
and the giant clump (GC) centroid in the instrumental CMD, and $(V-I, I)_{0,\rm GC} = (1.06, 14.34)$ 
is the calibrated position of the GC \citep{bensby13,nataf13}. From this, we find 
$(V-I, I)_{0,\rm S} = (1.15 \pm 0.07, 20.61 \pm 0.09)$, indicating that the source is a K-type 
main-sequence star.

Once the source type is known, we determine $\theta_{*}$ using the $VIK$ color-color relation 
\citep{bessell88} and the color-surface brightness relation \citep{kervella04}. 
The estimated angular size of the source is 
\begin{equation}
\theta_{*} = 0.390 \pm 0.032~{\mu}{\rm as},
\label{eq5}
\end{equation}
which corresponds to the angular Einstein radius  
\begin{equation}
\theta_{\rm E} = 0.065 \pm 0.009~{\rm mas}.
\label{eq6}
\end{equation}
With the measured Einstein timescale $t_{\rm E}$, the geocentric proper 
motion of the source relative to the lens is then 
\begin{equation}
\mu = {\theta_{\rm E} \over t_{\rm E}} = 3.16 \pm 0.46~{\rm mas~yr^{-1}}.
\label{eq7}
\end{equation} 
We note that the error in $\theta_{*}$ is derived from 
the uncertainty of the source brightness ($4\%$) 
and the color-surface brightness conversion ($7\%$). 
The errors of $\theta_{\rm E}$ and $\mu$ are then estimated based on
$\sigma_{\theta_{*}}$ and $\sigma_{\rho_{*}}$ (see Table~\ref{table:two}).   

The Einstein radius is related to the total mass of the lens, $M_{\rm tot}$, 
and the distance to the lens, $D_{\rm L}$, by
\begin{equation}
\theta_{\rm E} = \sqrt{{\kappa}M_{\rm tot}\pi_{\rm rel}}  \\
\sim 0.7 {\rm mas} \left({M_{\rm tot} \over 0.7~M_{\odot}}\right)^{1/2} \left({8~{\rm kpc} \over D_{\rm S}}\right)^{1/2} \left({{1-x} \over x}\right)^{1/2},
\label{eq8}
\end{equation}
where $\kappa = 4G/(c^{2}{\rm AU})$, $\pi_{\rm rel} = {\rm AU}(D_{\rm L}^{-1} - D_{\rm S}^{-1})$, $D_{\rm S}$ 
is the distance to the source, and $x = D_{\rm L}/D_{\rm S}$. 
Then, the derived $\theta_{\rm E}$ is significantly smaller than $\theta_{\rm E} \sim 0.7~{\rm mas}$ 
of an event caused by a stellar object with $M_{\rm tot} \sim 0.7~M_{\odot}$ and $x \sim 0.5$. 
This suggests that the event was generated by a VLM binary and/or 
a lens system located in the Galactic bulge very close to the source, i.e., $x \sim 1.0$.

\section{Physical Parameters}

The direct measurement of $M_{\rm tot}$ and $D_{\rm L}$ of the lens system requires the 
simultaneous detection of $\theta_{\rm E}$ and the microlens parallax $\pi_{\rm E}$, i.e., 
\begin{equation}
M_{\rm tot} = {\theta_{\rm E} \over \kappa\pi_{\rm E}};~~~~~D_{\rm L} = {{\rm AU} \over \pi_{\rm E}\theta_{\rm E} + \pi_{\rm S}},
\label{eq9}
\end{equation}
where $\pi_{\rm S} = {\rm AU}/D_{\rm S}$ is the source parallax \citep{gould92,gould04}. 
For OGLE-2017-BLG-1522, the angular Einstein radius is measured, but the microlens parallax 
cannot be measured, and thus the physical properties cannot be directly determined. 
Therefore, we investigate the probability distributions of physical parameters from 
the measured $\theta_{\rm E}$ and $t_{\rm E}$. For this, we perform a Bayesian analysis 
using a Galactic model based on the velocity distribution (VD), mass function (MF), and 
matter density profile (DP) of the Milky Way Galaxy.

In order to construct the model, we define the Cartesian coordinates in the Galactic frame
so that the center of the coordinates is the Galactic center and the $x$-axis and the $z$-axis point
toward the Earth and the north Galactic pole, respectively. Then, the line of sight distance $D$ of an 
object is related to the Galactic coordinates by
\begin{equation}
x = R_{\odot} - D{\rm cos}(l){\rm cos}(b),~y = D{\rm sin}(l){\rm cos}(b),~z = D{\rm sin}(b),
\label{eq10}
\end{equation}
where $R_{\odot} = 8~{\rm kpc}$ is the adopted Galactocentric distance of the Sun \citep{reid93,gillessen13}.

For the VD, we adopt the \citet{han95} model of $f(v_{y}, v_{z}) = f(v_{y})f(v_{z})$ 
which follows a Gaussian form, i.e., 
\begin{equation}
f(v_{y}) = {1 \over \sqrt{2\pi{\sigma_{y}}^{2}}} {\rm exp}\left[-{(v_{y} - \bar{v}_{y})^{2} \over 2{\sigma_{y}}^{2}}\right]
\label{eq11}
\end{equation}
and a similar form for $f(v_{z})$. Here $\bar{v}$ and $\sigma$ denote the mean and dispersion 
of the velocity component, respectively. Following \citet{han95}, we use 
$(\bar{v}_{z,\rm bulge}, \bar{v}_{z,\rm disk}) = (0, 0)~{\rm km}~{\rm s}^{-1}$ and
$(\sigma_{z,\rm bulge}, \sigma_{z,\rm disk}) = (100, 20)~{\rm km}~{\rm s}^{-1}$ for the $z$-direction and
$(\bar{v}_{y,\rm bulge}, \bar{v}_{y,\rm disk}) = (0, 220)~{\rm km}~{\rm s}^{-1}$ and
$(\sigma_{y,\rm bulge}, \sigma_{y,\rm disk}) = (100, 30)~{\rm km}~{\rm s}^{-1}$ for the $y$-direction. 
For a given set of the projected velocity of the lens $\bold{v}_{\rm L}$, the source $\bold{v}_{\rm S}$, 
and the observer $\bold{v}_{\rm o}$, the transverse velocity of the lens with respect to the source $\bold{v}$ 
is then given by
\begin{equation}
\bold{v} = \bold{v}_{\rm L} - \left[\bold{v}_{\rm S}{D_{\rm L} \over D_{\rm S}} + \bold{v}_{\rm o}{{D_{\rm S} - D_{\rm L}} \over D_{\rm S}}\right].
\label{eq12}
\end{equation}
The projected velocity of the observer $\bold{v}_{\rm o}$ is estimated by converting 
the heliocentric Earth's velocity at the peak time of the event, 
$\bold{v}_{\rm E} = (v_{\rm E, \rm E}, v_{\rm E, \rm N}) = (21.07, -1.75)~{\rm km}~{\rm s}^{-1}$, 
to the Galactic frame. In this procedure, we also consider the peculiar and circular motion 
of the Sun with respect to a local standard of rest, i.e., 
$\bold{v}_{\odot} = \bold{v}_{\odot,\rm pec} + \bold{v}_{\odot,\rm cir} = (12, 7) + (220, 0)~{\rm km}~{\rm s}^{-1}$.

For the MF, we separately consider the bulge and the disk lens populations. 
We model the bulge population by adopting the log-normal initial mass function 
\citep[IMF:][]{chabrier03}, while we model the disk population by adopting 
the log-normal present-day mass function \citep[PDMF:][]{chabrier03}. 
In both functions, the lower mass limit is set to $0.01~M_{\odot}$. 
We note that we do not consider stellar remnants, since planets 
orbiting a star in the phase of asymptotic giant branch or 
planetary nebular would be hard to survive, and thus the probability 
to find planets orbiting a remnant would be extremely low \citep[e.g.,][]{kilic09}.

For the DP, we use a triaxial profile for the bulge and a double exponential 
profile for the disk. In the case of bulge profile, we adopt the refined 
\citet{han03} model for which the profile follows the \citet{dwek95} model and 
the density is normalized by the star count results of \citet{holtzman98}. 
The disk profile is modeled by the two (thin and thick disk) 
exponential disks of the form
\begin{equation}
\rho_{r,z} = \rho(R_{\odot},0)e^{R_{\odot}/{h_{R}}}{\rm exp}\left(-{r \over h_{R}} - {{z + z_{\odot}} \over h_{Z}}\right),
\label{eq13}
\end{equation}
where $r = (x^{2}+y^{2})^{1/2}$, $z_{\odot}$ is the offset of the Sun from the Galactic plane, 
and $h_{R}$ and $h_{Z}$ are the radial and vertical scale length, respectively. 
We adopt $z_{\odot} = 25~{\rm pc}$, $(h_{R}, h_{Z})_{\rm thin} = (2600, 300)~{\rm pc}$, 
$(h_{R}, h_{Z})_{\rm thick} = (3600, 900)~{\rm pc}$, and the normalization factor 
$f = \rho_{\rm thick}(R_{\odot},0) / \rho_{\rm thin}(R_{\odot},0) = 0.12$ 
from \citet{juric08}, where they construct the profile using stellar objects 
detected by the Sloan Digital Sky Survey \citep[SDSS:][]{york00}. Finally, 
we normalize the number density to $\rho(R_{\odot},0) = 0.05~M_{\odot}~{\rm pc}^{-3}$ 
by following the disk density of solar neighborhood estimated by \citet{han03}.


With the adopted models of VD, MF, and DP, we generate microlensing events 
from the Monte Carlo simulation and then investigate the probability distributions of the host mass 
$M_{1}$ and lens distance using the measured $\theta_{\rm E}$ and $t_{\rm E}$ as a prior 
\footnote{We note that we apply the bulge model for the source by assuming that the source is 
in the bulge, while we separately apply the bulge and disk models for the lens. 
We then estimate the total probabilities by summing each set of the distribution.}. 
In Figure~\ref{fig:five}, we show the posterior probabilities of $M_{1}$ (upper panel) 
and $D_{\rm L}$ (lower panel) derived from our Bayesian analysis. 
The median value of the mass and the distance with $68\%$ $(1\sigma)$ confidence intervals are   
\begin{equation}
M_{1} = {0.045}_{-0.024}^{+0.076}~M_{\odot},~~~~~D_{\rm L} = {7.49}_{-0.88}^{+0.91}~{\rm kpc},  
\label{eq14}
\end{equation}
respectively. The estimated host star corresponds to a brown dwarf. 
The median value of the lens distance implies that both the lens and the source are 
likely to be located in the Galactic bulge. These consequently indicate that 
the extraordinarily small value of $\theta_{\rm E}$ is due to the 
combination of the small lens mass and the lens location close to the source.

From the measured $q$, the companion mass is determined by        
\begin{equation}
M_{2} = qM_{1} = 0.75_{-0.40}^{+1.26}~M_{\rm J}, 
\label{eq15}
\end{equation}
which roughly corresponds to the Jupiter mass planet. 
The physical companion-host projected separation is  
\begin{equation}
a_{\bot} = s{D_{\rm L}}{\theta_{\rm E}} = 0.59_{-0.11}^{+0.12}~{\rm AU}.
\label{eq16}
\end{equation}
By adopting that the snow line scales with the host mass 
\citep{kennedy08}, we estimate the snow line of the lens system by 
$a_{sl} = 2.7~{\rm AU}(M/M_{\odot}) \sim 0.12~{\rm AU}$, indicating that 
the giant plant is placed beyond the snow line.

\section{Discussion} 

\begin{figure}[th]
\epsscale{0.7}
\plotone{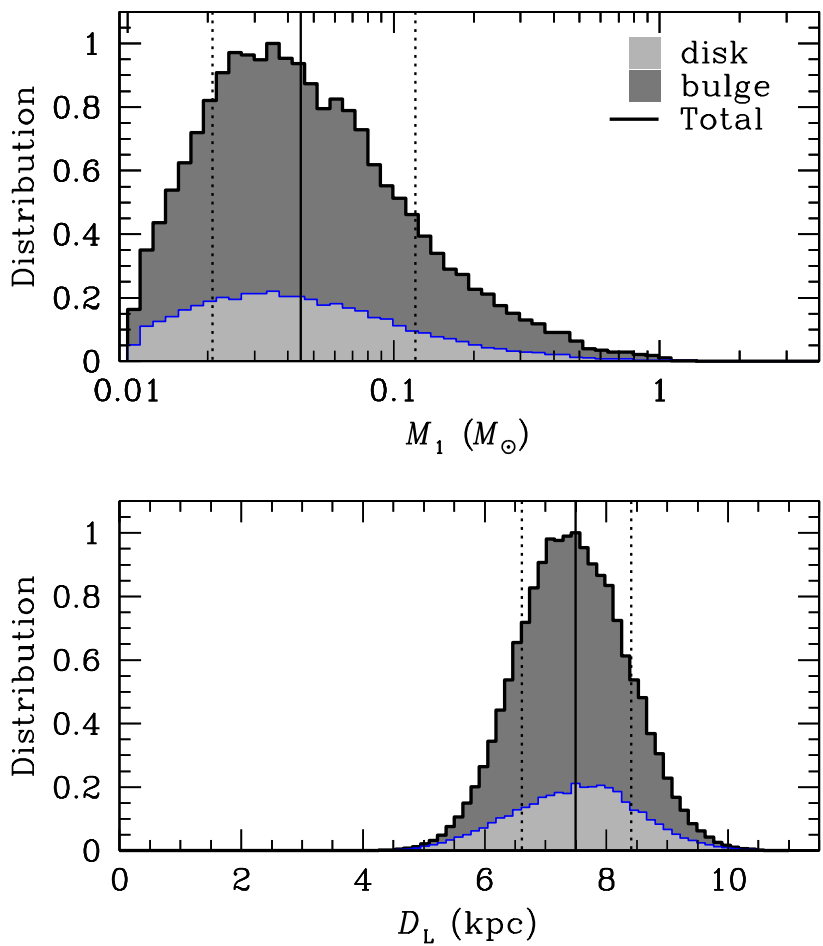}
\caption{\label{fig:five}
Probability distributions of the host mass (upper panel) and lens 
distance (lower panel) derived from the Bayesian analysis. 
The grey and darkgrey distributions are those derived from the disk and bulge 
Galactic models, respectively. In each panel, the solid line represents 
the median value and the two dotted lines represent the confidence 
intervals estimated based on the lower and upper boundaries 
encompassing the $68\%$ $(1\sigma)$ range of the distribution.
}
\end{figure}

\begin{figure}[th]
\epsscale{0.7}
\plotone{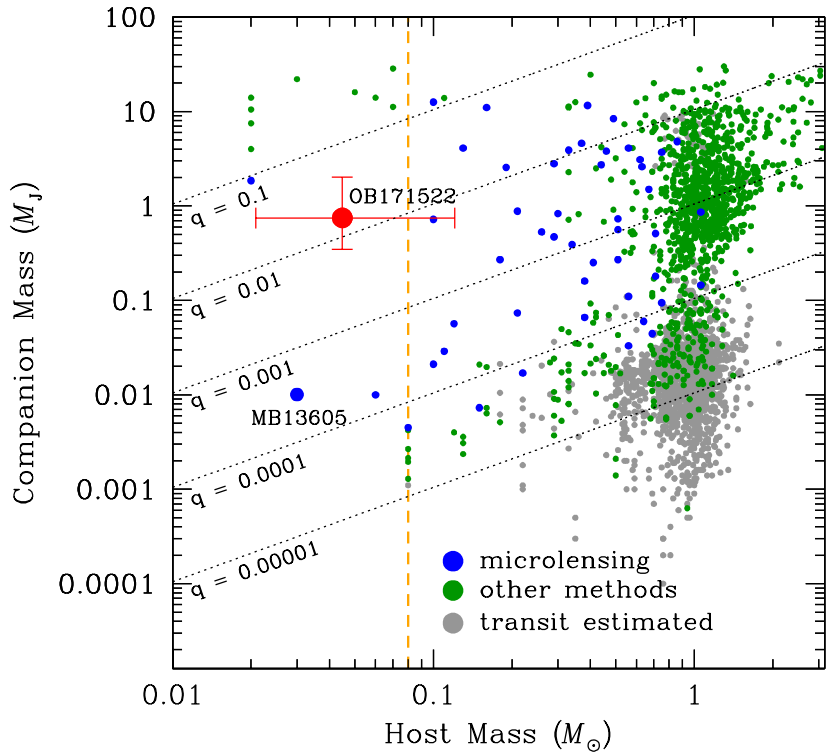}
\caption{\label{fig:six}
Mass distribution of known exoplanets. The red dot is OGLE-2017-BLG-1522. 
The planets discovered by microlensing method are marked by blue dots, while those 
found by other methods are marked by green dots. For transiting systems for which 
companion masses are not directly determined, we estimate the companion mass using 
the forecasting model of \citet{chen17} (grey dots). The yellow vertical dashed line 
represents the conventional star/brown-dwarf boundary. The values are acquired from 
http://exoplanetarchive.ipac.caltech.edu.
}
\end{figure}

We found a planetary candidate around a probable brown dwarf host 
located in the Galactic bulge. The probability that the host is a brown dwarf 
(i.e., $M_{1} < 0.08~M_{\odot}$) is $\sim$ 76\% (see Figure~\ref{fig:five}).
The light curve perturbations were 
clearly identified by the OGLE survey and continuously covered by 
the KMTNet observations despite the relatively short event timescale 
$(t_{\rm E} \sim 7.5~{\rm days})$, which enabled the unambiguous 
characterization of the lens system. This proves the capability of 
current microlensing experiments. In Figure~\ref{fig:six}, we compare 
the mass distribution of the lens to those of known exoplanets. 
From this, we find that this event could be the first binary that closes 
the gap between terrestrial and jovian mass companions in the brown-dwarf 
host region. We also find that MOA-2013-BLG-605 could have similar properties 
to our results \citep{sumi16}. However, it not only suffers from large uncertainties 
($3 \sim 21~M_{\oplus}$ for the companion and $0.03 \sim 0.2~M_{\odot}$ for the host) 
due to the severe degeneracy in the parallax measurement but also favors the terrestrial 
planet interpretation. It is worth noting that the analysis of 
a large sample of short-timescale binary events found by \citet{mroz17} can bring
additional information on the population of planetary-mass companions to
brown dwarf hosts.

Up to now, several brown dwarfs hosting giant planet companions have been discovered. 
However, their large mass ratios $(q \gtrsim 0.1)$ between the host and the companion 
would suggest that they are formed as binary systems either by the dynamical interaction in 
unstable molecular clouds \citep{bate09,bate12} or by the turbulent fragmentation 
of molecular cloud cores \citep{padoan04}. This implies that the companions could be 
considered as substellar objects rather than planets \citep{chabrier14}. By contrast, 
the low mass ratio of this event $(q \sim 0.016)$, combined with recent reports of 
the massive disks $(\gtrsim M_{\rm J})$ around young brown dwarfs 
\citep{hervey12,andre12,palau14}, suggest that the planetary companion of 
the event may be formed by planet formation mechanisms. Therefore, 
OGLE-2017-BLG-1522Lb could be the first giant planet orbiting 
around a brown-dwarf host having a planetary mass ratio.

It would therefore be of considerable interest to make a definitive determination 
of whether the host is a brown dwarf or a star. Considering that the measured 
relative source motion is $\mu \sim 3~{\rm mas~yr^{-1}}$, the source will be separated 
from the lens about $\gtrsim 30~{\rm mas}$ at first light of the upcoming next generation 
($D \sim 30$m class) telescopes. Then, the source and the lens could be resolved with 
these telescopes because their resolution in $H$ band will be 
$\theta = 14(D/30{\rm m})^{-1}~{\rm mas}$. Hence, at first light for any of these 
telescopes, the source and lens will be separated by well over $1$ FWHM. If the host 
is a star rather than a brown dwarf, $M > 0.08~M_\odot$, then $M_{H,{\rm host}} \la 9$ 
and $\pi_{\rm rel} < 6.4~{\mu}{\rm as}$, i.e., $D_{\rm L} \sim 8~{\rm kpc}$. The $H$ band 
magnitude of the host is then $H_{{\rm host}} \la 23.5$ which should be easily visible 
in AO images. As a result, the nature of the host (star or brown dwarf) can be unambiguously 
resolved at that time.

\acknowledgments
This research has made use of the KMTNet system operated by the Korea 
Astronomy and Space Science Institute (KASI) and the data were obtained at 
three host sites of CTIO in Chile, SAAO in South Africa, and SSO in Australia. 
OGLE project has received funding from the National Science Centre, Poland, 
grant MAESTRO 2014/14/A/ST9/00121 to AU. C. Han was supported by grant 
2017R1A4A1015178 of the National Research Foundation of Korea. Work by 
WZ, YKJ, and AG were supported by AST-1516842 from the US NSF. WZ, IGS, 
and AG were supported by JPL grant 1500811. AG is supported from KASI 
grant 2016-1-832-01.

\begin{deluxetable}{lccc}
\tablecaption{Error Correction Parameters\label{table:one}}
\tablewidth{0pt}
\tablehead{
\multicolumn{1}{c}{Observatory}  &
\multicolumn{1}{c}{Number}       &
\multicolumn{1}{c}{$k$}          &
\multicolumn{1}{c}{$\sigma_{\rm min}$ (mag)}
}
\startdata
OGLE $(I)$        & 11365  &  1.672   & 0.002  \\
KMTC BLG03 $(I)$  &  1336  &  1.153   & 0.000  \\
KMTC BLG43 $(I)$  &  1639  &  1.438   & 0.000  \\
KMTS BLG03 $(I)$  &  2323  &  1.268   & 0.000  \\
KMTS BLG43 $(I)$  &  2173  &  1.389   & 0.000  \\
KMTA BLG03 $(I)$  &  1787  &  1.679   & 0.000  \\
KMTA BLG43 $(I)$  &  1818  &  1.506   & 0.000  
\enddata
\end{deluxetable}

\begin{deluxetable}{lr}
\tablecaption{Lensing Parameters\label{table:two}}
\tablewidth{0pt}
\tablehead{
\multicolumn{1}{l}{Parameters} &
\multicolumn{1}{c}{Values} 
}
\startdata
$\chi^2$/dof                 &   21009.3/22434      \\
$t_{0}$ (${\rm HJD'}$)       &  7971.80 $\pm$ 0.01  \\
$u_{0}$ ($10^{-2}$)          &     5.39 $\pm$ 0.34  \\
$t_{\rm E}$ (days)           &     7.53 $\pm$ 0.28  \\
$s$                          &     1.21 $\pm$ 0.01  \\
$q$ ($10^{-2}$)              &     1.59 $\pm$ 0.16  \\
$\alpha$ (rad)               &     6.22 $\pm$ 0.01  \\
$\rho_\ast$ ($10^{-2}$)      &     0.60 $\pm$ 0.07  \\                     
$f_{\rm S}$                  &    0.035 $\pm$ 0.002 \\
$f_{\rm B}$                  &    2.221 $\pm$ 0.002
\enddata 
\vspace{0.05cm}
\tablecomments{
${\rm HJD}'= {\rm HJD}-2,450,000~{\rm days}$
}
\end{deluxetable}

\end{document}